# SVM Model for Identification of human GPCRs

Sonal Shrivastava, K. R. Pardasani, M. M. Malik

**Abstract**— G-protein coupled receptors (GPCRs) constitute a broad class of cell-surface receptors in eukaryotes and they possess seven transmembrane α-helical domains. GPCRs are usually classified into several functionally distinct families that play a key role in cellular signalling and regulation of basic physiological processes. We can develop statistical models based on these common features that can be used to classify proteins, to predict new members, and to study the sequence–function relationship of this protein function group. In this study, SVM based classification model has been developed for the identification of human gpcr sequences. Sequences of Level 1 subfamilies of Class A rhodopsin are considered as case study. In the present study, an attempt has been made to classify GPCRs on the basis of species. The present study classifies human gpcr sequences with rest of the species available in GPCRDB. Classification is based on specific information derived from the n-terminal and extracellular loops of the sequences, some physicochemical properties and amino acid composition of corresponding gpcr sequences. Our method classifies Level 1 subfamilies of GPCRs with 94% accuracy.

—————————— ◆ ——————————

## 1 INTRODUCTION

G-protein coupled receptors (GPCRs) are one of the largest superfamilies of membrane proteins in mammals. They are one of the most targeted protein families in pharmaceutical research today as they play important roles in a variety of biological and pathological processes such as development and proliferation, neuromodulation, angiogenesis, metabolic disorders, inflammation, and viral infection [2], [4], [8]. More than one third of the drugs in the market acts on GPCRs[6], [10]. One of the problems with GPCRs is that they may lack enough primary sequence similarities even though they share similar structures, biochemical properties and functions. Therefore in order to classify these proteins, classifier should consider various other parameters except relying solely on alignments. Since the physicochemical properties of proteins have enough specific information, so they can be used to identify proteins that share similar functions even from short and diverged sequences where alignments cannot be reliable [7].

GPCRs can be grouped into six distinct classes A,B,C,D,E and frizzled/Smoothened family as defined by GPCRDB classification on the base of shared sequence motifs [15]. Traditionally, though, GPCRs' family classification is based on receptor's ligand specificity. However, common ligand specificity does not necessarily infer a certain level of sequence identity. Somehow some higher-order relationship between sequence and binding of ligands of a particular chemical class seems to exist, as revealed by phylogenetic analysis.

All GPCRs share a common structural feature – seven helical transmembrane region(7TM) that anchors the receptor to the plasma membrane of the cell. The N-terminal is exposed to the extracellular space whereas the C-terminal is extended in the cytoplasm. These proteins transduce an extracellular signal into the cell via a guanine binding protein (G-protein) on activation upon ligand binding at extracellular space. Now this signal plays an important role in the regulation of a large number of metabolic processes such as neurotransmission, hormonal secretion, cellular differentiation and metabolism [14]. Thus functional analysis of GPCRs is very important for understanding the signal transduction processes.

GPCR is an essential subject of many recent biomolecular projects. They are responsible for diverse physiological processes such as neurotransmission, secretion, cellular metabolism growth and cellular differentiation as well as inflammatory and immune responses. Therefore, they are vital for the research and development for new drugs [6]. Various databases have been created in order to observe and categorize different characteristics of GPCRs. These databases hold sequences, mutation data and ligand binding data. Moreover, these databases are further improved by multiple sequence alignments, two dimensional visualization tools, three dimensional models and phylogenetic trees [10]. Even though thousands of GPCR sequences are known as a result of ongoing genomics projects, the crystal structure has been discovered only for one GPCR sequence using electron diffraction at medium resolution (2.8 A) to date and for many of the GPCRs the activating ligand is unknown, which are called orphan GPCRs [7]. Hence, based on sequence information, a functional classification method of those orphan GPCRs and new upcoming GPCR sequences is crucial to identify and characterize novel GPCRs.

In this study, primary sequences of n-terminal and extracellular loops are considered rather than the whole sequence of GPCR proteins. Since the families of GPCRs are closely correlated with their structure and function, it is significant to develop an automatic computational method for identifying these transmembrane proteins from genome sequences and classifying them based on their biological function.

————————————

- *Sonal Shrivastava is with the Maulana Azad National Institute of TechnologyBhopal, India.*
  - *K.R. Pardasani is with the Department of Mathematics, Computer Applications and Bioinformatics, MANIT, Bhopal, India.*
- *M.M.Malik is with the Physics Department, MANIT, Bhopal, India.*

In the current literature, to classify GPCRs in different levels of families, there exist different methods like prim database search tools, e.g., BLAST, FASTA. However, these methods only work if the query protein sequence is highly similar to the existing database sequences in order to work properly [7]. In addition to these database search tools, the classification of GPCRs is also done by using Hidden Markov Models [16], bagging classification trees [11] and SVMs [3], [14], [17]. In a recent work Bakir et al. used a fixed length feature vector of 40 most distinguishing patterns to classify amine sub-family GPCRs with 97% accuracy using SVMs. But in all the above mentioned work, classification is based on functional or structural basis and accordingly all the gpcr sequences are divided in different families and subfamilies. But no computational approach is reported in literature for classification of gpcrs based of species. Here an attempt has been made to distinguish the gpcr sequences found in homo sapiens from rest of the species. An SVM model has been developed for classification incorporating the ligand specific properties of GPCRs.

## 2    Material & Methods

### 2.1  Amino Acid Sequence Data for G-Protein Couple Receptors

All the GPCRs were extracted from the GPCRDB information system March 2005 release 9.0 [10]. GPCRDB is a molecular class-specific information system that collects heterogeneous data on GPCRs, including data on sequences, ligand-binding constants and mutations.

All protein sequences in GPCRDB are mainly divided into six families: rhodopsin-like family, secretin-like family, metabotropic glutamate family, fungal pheromone family, cAMP receptor family and frizzled/smoothened family. Each family covers several subfamilies of GPCRs. In addition, some putative families, orphan receptors and non-GPCRs families are also included in the database. Detailed description refers to the website via http://www.gpcr.org/7tm/ [10].

The GPCRDB is used to extract the sequences of subfamilies of class A rhodopsin family. In this study we have selected five subfamilies of class A rhodopsin family – amine, olfactory, peptide, rhodopsin and prostanoid.

### 2.2  Extraction of required data from amino acid sequences of GPCRs

TMHMM server 2.0 is used to extract the intracellular and extracellular loops and TM region. A TMHMM prediction service is available at http://www.cbs.dtu.dk/services/TMHMM/ (Copyright 2001 Academic Press).

It is found in data analysis that C-terminal loops (CTL) of GPCRs have shown clear variety in length as well as in physical properties even within the same GPCR families. Thus they cannot be treated as suitable parameters for classification of GPCRs. Finally, we've selected extracellular loop lengths and the n-terminal loop length for classification.

### 2.3  The physicochemical properties of GPCRs

Web-server COPid is used for annotating the function of a protein from its composition using whole or part of the protein. COPid has three modules called search, composition and analysis [13]. The search module allows searching of protein sequences in six different databases. Search results list database proteins in ascending order of Euclidean distance or descending order of compositional similarity with the query sequence. The composition module allows calculation of the composition of a sequence and average composition of a group of sequences. The composition module also allows computing composition of various types of amino acids like (e. g. charge, polar, hydrophobic residues). The analysis module provides the following options; i) comparing composition of two classes of proteins, ii) creating a phylogenetic tree based on the composition and iii) generating input patterns for machine learning techniques. The COPid web-server is available at http://www.imtech.res.in/raghava/copid/.

We have used COPid to determine the composition of amino acids in all the gpcr sequences.

*Amino Acid Composition*—Protein information can be encapsulated in a vector of 20 dimensions, using amino acid composition of the protein. In the past, this approach has been used for predicting the subcellular localization of proteins [5], [13]. The amino acid composition is the fraction of each amino acid type within a protein. The fractions of all 20 natural amino acids were calculated by using Equation 1,

$$\text{Fraction of aa}i = \frac{\text{total number of amino acids of type i}}{\text{total number of amino acids in protein}}$$

(Eq. 1)

where i is an specific type of amino acid (aa).

### 2.4 Software and Algorithm for Classification

The Weka Data Mining Java script 3.7 was used for training and testing the corresponding data set and for the comparison with other learning algorithm (Obtained through the Internet: http://www.cs.waikato.ac.nz/~ml/weka/ ). All the algorithms used were taken from the Weka suite. Support vector machines are used for classification.

## 3    Results and Discussion

Support vector machines (SVMs) are a set of related supervised learning methods used for classification and regression [9]. Viewing input data as two sets of vectors in an $n$-dimensional space, an SVM will construct a separating hyper plane in that space, one which maximizes the margin between the two data sets. To calculate the margin, two parallel hyperplanes are constructed, one on each side of the separating hyper plane, which are "pushed up against" the two data sets. Intuitively, a good separation is achieved by the hyperplane that has the largest distance to the neighbouring data points of both classes, since in general the larger the margin the lower the generalization error of the classifier [9], [12].

Not all attributes, however, are fit for use in a classifier. First, some attributes are clearly not independent and do not provide any additional advantage when evaluated together. Thus after determining the sequence lengths of n-terminal, extracellular, intracellular loops and TM regions, some sequences are filtered out because of missing values. Many parameters, such as molecular weight, are discarded as they are not contributing for classification purpose. As we planned this work to focus on ligand specific properties, the intracellular loops were discarded. Finally we had 224 instances with their n-terminal and extracellular loop lengths. After data filtering, 188 sequences are used as the training data whereas rest of the 36 sequences are used as test data.

The input for the SVM is a fixed-length vector obtained using amino acid composition from the primary amino acid sequence. Three extracellular loops of the amino acid sequences along with their n-terminus loops are taken as parameters from five different subfamilies of class A, rhodopsin receptors. Our SVM based model for classification of subfamily of rhodopsin is able to classify the level 1 subfamily human receptors with 94.44% accuracy.

The performance of the SVM module was validated using a cross-validation test. The performance of the module developed for discriminating between human GPCRs and other gpcr sequences is summarized in Table 1. The results depict that the method can differentiate human GPCRs from other gpcr sequences with an accuracy of 94.44%. The best results were obtained using the RBF kernel with $\gamma$ = 10. The value of regulatory parameter C was optimized to 1.0. Training is done using SVM available in weka suite 3.7.

To automate the process, computational methods such as decision trees, naive bayes, neural networks and support vector machines (SVMs), have been extensively used in the fields of classification of biological data. Among these methods, SVMs give best prediction performance, when applied to many real-life classification problems, including biological issues. One of the most critical issues in classification is the minimization of the probability of error on test data using the trained classifier, which is also known as structural risk minimization. It has been demonstrated that SVMs are able to minimize the structural risk through finding a unique hyper-plane with maximum margin to separate data from two classes. Therefore, compared with the other classification methods, SVM classifiers supply the best generalization ability on unseen data [3]. This table clearly explains our reason to choose SVM as compared to other classification algorithms. MCC in case of SVM comes out to be **0.89**.

The statistics is given below:

| Method | Sensitivity | Specificity | Accuracy |
|---|---|---|---|
| BFT | 65.38 | 69.16 | 67.41 |
| J48 | 67.85 | 74.10 | 70.98 |
| Baye's Net | 64.58 | 66.40 | 65.62 |
| Naive Bayes | 54.83 | 58.77 | 57.14 |
| SVM | 100.00 | 90.90 | 94.44 |

Table 1

| | | |
|---|---|---|
| Correctly Classified Instances | 34 | 94.4444 % |
| Incorrectly Classified Instances | 2 | 5.5556 % |
| Kappa statistic | 0.8861 | |
| Mean absolute error | 0.0556 | |
| Root mean squared error | 0.2357 | |
| Relative absolute error | 11.214 % | |
| Root relative squared error | 47.4146 % | |
| Total Number of Instances | 36 | |

## 4    Conclusion

We have worked on a new idea of classifying human GPCRs from other gpcr sequences based upon sequence characteristics. Till date no such classification system is available. The establishment of such methods can help in detection of an unknown GPCR sequence as human sequence which may prove useful in forensic sciences. Classification of GPCRs based on species can be of utmost importance in deriving the characteristics and signature motifs for different species, which can further be useful in case studies like animal meat production, agrochemical industries, phylogenetic studies or in identification of human specific diseases. Such models can be developed to identify the specific characteristics of human gpcrs as compared to other species, which can be useful in drug discovery.

**Sonal Shrivastava** is a research scholar in department of Bioinformatics from Maulana Azad National Institute of Technology, Bhopal. She has completed her M.Sc. in Computer Science from Jiwaji University, Gwalior. Her current research interests are in the field of pattern recognition, machine learning techniques and computational biology.

**Prof. K. R. Pardasani** is the head of Department of Mathematics, Bioinformatics and Computer Applications. Obtained Ph.D. in area of Bio-Computing in 1988 from School of Mathematics and Allied Sciences,Jiwaji University,Gwalior. Worked under the Supervision of Dr.V.P.Saxena. Title of thesis is "Mathematical Investigations on Human Physiological Heat Flow Problems with Special Relevance to Cancerous Tumors". Passed M.Sc. (Pure Mathematics) in 1984 with first division (71%) from SOS in Mathematics, Jiwaji University, Gwalior and stood third in Merit List of the University. He has more than 125 publications in various national and international journals.

**Dr. M. M. Malik** is Asst. Professor, deptt. Of Physics. He has published 19 research papers in various reputed journals.